\documentclass[prd,nofootinbib,english]{revtex4-2}
\usepackage[utf8]{inputenc}

\usepackage{graphicx,color} 
\usepackage{xcolor}
\usepackage{rotating}
\usepackage{amssymb}
\usepackage{soul}
\usepackage{amsfonts}
\usepackage{amsmath}
\usepackage{xspace}
\usepackage{mathtools} 
\usepackage{verbatim}
\usepackage{comment}
\usepackage{enumitem}
\usepackage{physics}
\usepackage{graphicx}
\usepackage[colorlinks]{hyperref}
\usepackage{MnSymbol}
\usepackage{graphicx}
\usepackage{accents}
\usepackage[normalem]{ulem}

\newcommand\ie{\mbox{\textit{i.\,e.}}\xspace}

\newcommand\eg{\mbox{e.\,g.}\xspace}

\newcommand{\Ntot}{\vec N_{\rm tot}}
\newcommand{\ptot}{\vec p_{\rm tot}}

\newcommand{\Rtot}{\vec R_{\rm tot}}

\begin{document}

\title{Universality of free  fall in Planck-scale deformed Newtonian gravity}

\author{Giuseppe Fabiano}
\email{gfabiano@lbl.gov}
\affiliation{Physics Division, Lawrence Berkeley National Laboratory, Berkeley, CA}\affiliation{Department of Physics, University of California, Berkeley, CA 94720, USA}\affiliation{Centro Ricerche Enrico Fermi—Museo Storico della Fisica e Centro Studi e Ricerche “Enrico Fermi”, Roma}

\author{Domenico Frattulillo}
\email{domenico.frattulillo@na.infn.it}
\affiliation{INFN, Sezione di Napoli, Complesso Univ. Monte S. Angelo, I-80126 Napoli, Italy}

\author{Christian Pfeifer}
\email{christian.pfeifer@zarm.uni-bremen.de}
\affiliation{ZARM, University of Bremen, 28359 Bremen, Germany.}

\author{Fabian Wagner}
\email{f.wagner@thphys.uni-heidelberg.de}
\affiliation{Institut f\"ur theoretische Physik, Universit\"at Heidelberg, Philosophenweg 16, 69120
Heidelberg, Germany}

\begin{abstract}
    The universality of free fall is one of the most cherished principles in classical gravity. Its fate in the quantum world is one of the   key questions in fundamental physics.
    We investigate the universality of free fall in the context of Planck scale modifications of Newtonian gravity. Starting from a doubly-special-relativity setting we take the Newtonian limit to obtain deformed Galilean relativity. We study the interaction between two test particles, subject to deformed Galilean relativity, and a classical, undeformed gravitational source, the Earth. Such an interaction is investigated here for the first time.  Considering the two test particles falling freely in the source's gravitational field, we examine whether the universality of free fall is affected by deformed relativistic symmetries. We  show that, in general, the universality of free fall is violated. Remarkably, we find that there exist distinguished models for which the universality of free fall is realized and which predict a specific modification of the Newtonian potential.
\end{abstract}

\maketitle

\tableofcontents

\section{Introduction}
The universality of free fall is a key property of general relativity . It played a crucial role in the development of Newton's as well as in Einstein's theory of gravity  as one of the many equivalent characterizations of the weak equivalence principle (WEP). Yet, the classical description of gravity is incompatible with the principles of quantum theory and  must eventually be replaced by a quantum theory of gravity (QG). An immediate question which arises is how the universality of free fall carries over to a theory of quantum gravity, and, whether it can help as a guideline to construct such a theory. In this article we  study and discuss aspects and partial answers to this question by means of phenomenological quantum-gravity models \cite{Addazi:2021xuf}. 

More precisely, we  work within the framework of doubly (or deformed) special relativity (DSR) \cite{Amelino-Camelia:2000stu,Amelino-Camelia:2002cqb,Magueijo:2001cr} in which the Planck energy scale, expected to be relevant for quantum gravity phenomena, is elevated to a relativistic invariant akin to the speed of light. Such models emerge, for example, from quantum deformations of the Poincar\'e algebra \cite{Kowalski-Glikman:2002iba}, non-commutative spacetime geometry \cite{Amelino-Camelia:1999jfz,Borowiec:2010yw}, 3D quantum gravity \cite{Matschull:1997du,Freidel:2003sp,Freidel:2005me,Freidel:2005bb,Cianfrani:2016ogm}, loop quantum gravity \cite{Bojowald:2012ux,Amelino-Camelia:2016gfx}, string theory \cite{Veneziano:1986zf,Kostelecky:1988zi,Yoneya:2000bt}. We understand DSR as a generic phenomenological model for a vacuum state of quantum gravity, thus incorporating a potential nonperturbative effect of quantum spacetime in a deformed classical flat or curved spacetime geometry \cite{Amelino-Camelia:1999iec,Addazi:2021xuf,Barcaroli:2015xda,Barcaroli:2017gvg,Pfeifer:2021tas}.

The Planck scale can be made invariant by deforming the Poincar\'e symmetries and  charge composition laws to preserve the relativity principle, bringing about nonlinearities in momentum space such as torsion or curvature \cite{Kowalski-Glikman:2002iba,Kowalski-Glikman:2002eyl}. This is in contrast with Lorentz Invariance Violation (LIV) models \cite{Colladay:1998fq,Kostelecky:2011qz}, where the Planck-scale effects introduce a preferred frame, by, for example, only modifying the dispersion relation of particles and not modifying the observer transformations from linear Lorentz transformations to deformed Lorentz transformations.

A central feature of DSR models is that deforming the generators of total translations makes locality a relative concept \cite{Amelino-Camelia:2011lvm,Amelino-Camelia:2011hjg,Gubitosi:2011hgc,Amelino-Camelia:2011uwb}. Whether an event, such as a scattering process, is local can only be determined by an observer at the event. In a reference frame translated from the event, the same process appears nonlocal. The relativity of locality is analogous to the relativity of simultaneity in the transition from Galilean to Lorentzian relativity.

In this paper, we aim to examine the implications of deformed spacetime symmetries for the universality of free fall. Specifically, we explore deformations of Newtonian gravity inspired by the Wigner-\.In\"on\"u contraction of the algebra sector of the $\kappa$-Poincar\'e Hopf algebra \cite{Brunkhorst:2017gcz}. Following an operational approach, we model Galileo's leaning tower of Pisa experiment, which leads us to a quantitative characterization of possible deviations from the universality of free fall in DSR. The approach to our investigation can be summarized as follows:

\begin{itemize}
    \item We apply the Wigner-\.In\"on\"u contraction to the algebra sector of the bicrossproduct $\kappa$-Poincar\'e algebra \cite{Majid:1994cy}, finding that the deformation scale ($\mu$) in the deformed Galilean-relativistic regime has to have units of mass. The absence of a velocity scale, which is typical in Galilean relativity, implies that deformations depend solely on the mass of the particles involved. This is remarkable because mass is a central element of the algebra and Poisson commutes with all generators. As a result, the Galilean counterpart of the $\kappa$-Poincar\'e algebra forms a Lie algebra for the single particle sector.
    
    \item We construct a general deformed dynamical model of two elementary particles, which obey deformed Galilean relativistic symmetries, interacting gravitationally with a macroscopic object (the Earth). Dynamical DSR models have rarely been studied in the literature \cite{Amelino-Camelia:2014gga,Bosso:2023nst,Amelino-Camelia:2023rkg}, and here we include gravity for the first time. 
    
    \item We model Galileo's tower of Pisa experiment, by allowing two distinct objects to fall from the same initial position, as seen by a local observer on top of the tower. Using our DSR model of gravitational interaction, we find that two objects released simultaneously in the frame on the top of the tower of Pisa, generally, do not reach an observer at the ground simultaneously.
    
    \item Remarkably, we find that it is possible to construct DSR models, by choosing the model parameters in a specific way, such that the universality of free fall is maintained. In order for the theory to be self-consistent these choices of parameters imply necessarily a modification of the Newtonian gravitational potential.

    This result extends and complements the LIV analysis of the effect of modified dispersion relations (MDRs) on the WEP presented in \cite{Hohmann:2024lys}, where a violation of the WEP was found, for all MDRs that do not satisfy a homogeneity scaling property with respect to the momenta.
\end{itemize}

We structure the presentation of our results as follows. In Section \ref{sec:contraction} we review the expressions for the algebra sector of the bicrossproduct $\kappa$-Poincar\'e algebra and obtain a deformed Galilean algebra\footnote{To be precise, we find a deformation of the Bargmann algebra, which is a central extension of the Galilean algebra adding the mass as central element.} as its Wigner-\.In\"on\"u contraction. In Section \ref{sec:DSRModel} we construct a deformed phenomenological model of two test particles interacting with a massive gravitating body. In Section \ref{sec:EotvosDSR} we discuss conditions for the universality of free fall as well as phenomenological prospects of our model. We conclude our discussion in Section \ref{sec:discussion} and check the robustness of our claims by increasing the number of free model parameters in Appendix \ref{app:MDR}. 

\section{Galilean contraction of the $\kappa$-Poincar\'e algebra\label{sec:contraction}}

 In this section, we obtain the deformed Galilean algebra underlying our DSR model by applying the Wigner-\.In\"on\"u contraction to the $\kappa$-Poincar\'e Hopf-Algebra in the "bicrossproduct basis" \cite{Majid:1994cy,Gubitosi:2011hgc}, see also \cite{Brunkhorst:2017gcz}. This sets the stage for the implementation of a DSR model of the gravitational interaction on Newtonian level in the next Section \ref{sec:DSRModel}.
 The relevant Poisson brackets among the symmetry generators for the deformed relativistic symmetry algebra read 
\begin{equation}
\label{eq:kappaalgebra}
\begin{aligned}
    \{P_\mu,P_\nu\}&=0\,,  & \{R_i,P_0\}&=0\,,\\
    \{R_i,P_j\}&=\epsilon_{ijk}P_k\,,  & \{R_i,R_j\}&=\epsilon_{ijk}R_k\, , \qquad \mu=0,1,2,3\\
    \{R_i,N_j\}&=\epsilon_{ijk}N_k\,, & \{N_i,P_0\}&=P_i\, , \qquad\quad\quad i=1,2,3 \\
   \{N_i,P_j\}&=\delta_{ij}\qty(\frac{\kappa}{2c^2}(1-e^{-2 P_0/\kappa})+\frac{1}{2\kappa}{\vec{P}}^2)- \frac{P_i P_j}{\kappa}\,,  & \{N_i,N_j\}&=-\frac{1}{c^2}\epsilon_{ijk}R_k,
    \end{aligned}
\end{equation}
where $P_\mu$ are the generators of translations, $R_i$ are the generators of rotations, $N_i$ are the generators of boosts and $\kappa$ is the model parameter with units of energy expected to be of the order of the Planck energy $E_{\rm P}=M_{\rm P}c^2$ ($M_{\rm P}$ being the Planck mass). The Casimir operator of the algebra is
\begin{equation}\label{eq:Casimir element}
C=\qty(2\kappa\sinh(\frac{P_0}{2\kappa}))^2-e^{P_0/\kappa}{\vec{P}}^2c^2.
\end{equation}
which in physical applications determines the dispersion relation of a particle of mass $m$ and energy-momentum $p_\mu$ by means of\footnote{As the model features an energy scale, in principle any function of the Casimir operator could generate a valid dispersion relation. We treat this case in Appendix \ref{app:MDR}.}
\begin{equation}\label{eq:kappadisprel}
    m^2c^4=\qty(2\kappa\sinh(\frac{p_0}{2\kappa}))^2-e^{p_0/\kappa}{\vec{p}}^2c^2.
\end{equation}
All algebraic relations reduce to the the Poincar\'e algebra for $\kappa \to \infty$.

The Casimir constraint \eqref{eq:kappadisprel} can be solved for $p_0$ to obtain the on-shell relation
\begin{equation}\label{eq:onshell}
   p_0= \kappa\log \left(\frac{2+\sqrt{(mc^2/\kappa)^4+4(mc^2/\kappa)^2+4 (pc/\kappa)^2}+ (mc^2/\kappa)^2}{2-2(pc/\kappa)^2}\right) \, .
\end{equation}

Following \cite{Arzano:2022nlo,Ballesteros:2020uxp}, the symmetry generators can be contracted to their Galilean limit by means of the rescaling $\kappa=\mu c^2$, where $\mu$ has dimensions of a mass, and taking the limit $c\rightarrow \infty$. 
 As a result, the non-vanishing Poisson brackets are
\begin{equation}
\label{eq:kappaalgebracontracted}
\begin{aligned}
    \{R_i,P_j\}&=\epsilon_{ijk}P_k\, ,  & \{R_i,R_j\}&=\epsilon_{ijk}R_k \, ,\\
    \{R_i,N_j\}&=\epsilon_{ijk}N_k \, , 
   & \{N_i,P_j\}&=\frac{\mu}{2}\left(1-F(\epsilon)^2\right)\delta_{ij},
    \end{aligned}
\end{equation}
where we introduced the function
\begin{equation}
    F(\epsilon)=\frac{1}{1+\frac{1}{2} \epsilon \left(\sqrt{4+\epsilon^2}+\epsilon\right)},
\end{equation}
and the dimensionless parameter $\epsilon=m/\mu$, which is expected to be small for elementary particles (recall that $\mu$ is of the order of the Planck mass). The on-shell relation \eqref{eq:onshell} can be expanded in $c$ as
\begin{equation}
    p_0(m,\vec{p})=-\frac{m}{\epsilon} c^2 \log F(\epsilon)+\frac{\abs{\vec{p}}^2}{2m}\frac{2}{\sqrt{4+\epsilon^2}F(\epsilon)}+O(1/c^2).
\end{equation}
Remarkably, the Wigner-\.In\"on\"u contraction of the generators of the $\kappa$-Poincar\'e Hopf-algebra provided here reduces the energy deformation scale $\kappa$ to the mass deformation scale $\mu$. As there is no scale with units of velocity left in the theory, by dimensional analysis, the deformations can only involve added factors proportional to powers of $m/\mu$. Given that the mass is a parameter and commutes with all generators, the algebra defined by the non-trivial Poisson brackets in \eqref{eq:kappaalgebracontracted} is a Lie-algebra, in contrast to the full algebra in \eqref{eq:kappaalgebra}. The contracted algebra reduces to the Galilei algebra in the  limit $\mu \to \infty$.

For phenomenological purposes, from now on we work at first order in $\epsilon$, at which the Poisson brackets between the conserved charges can be written as
\begin{equation}
\label{eq:algfirstorder}
\begin{aligned}
 \{R_i,P_j\}=&\epsilon_{ijk}P_k \,, &&\qquad\{R_i,R_j\}=\epsilon_{ijk}R_k\,,\\
    \{R_i,N_j\}=&\epsilon_{ijk}N_k\,,  
   &&\qquad\{N_i,P_j\}=\delta_{ij} \,m\qty(1-\epsilon) \, ,
\end{aligned}
\end{equation}
while the on-shell relation assumes the form
\begin{equation}
\label{eq:kinfirstorder}
p_0(m,\vec{p})=mc^2+\frac{\vec{p}^2}{2m}(1+\epsilon)\,,
\end{equation}
and leads to the definition of the free-particle Hamiltonian 
\begin{equation}
\label{eq:defkinenergy}
    H^{\rm free}=\frac{\vec p^2}{2m}(1+\epsilon) \, ,
\end{equation}
where we have neglected the constant rest-mass term.
Based on the analysis conducted in \cite[Eq. (63)]{Hohmann:2024lys}, this on-shell MDR violates the WEP, as long as $\epsilon \neq 0$. Below we see that, taking  into account the deformed momentum composition laws for two freely falling bodies in addition to the MDR, this must not necessarily be the case if one allows for modifications of the Newtonian potential.

Having introduced the Galilean contraction of the $\kappa$-Poincar\'e algebra, let us now construct a model of the gravitational interaction based on these symmetries.

\section{DSR Model of Gravitational interaction\label{sec:DSRModel}}

In this section, we construct a DSR model of two test-particles interacting gravitationally with the Earth. Here, the meaning of test-particles is the conventional one adopted in classical physics: the gravitational interaction among themselves is negligible with respect to that with the external field and they do not perturb the external gravitational field, which in our case is the one generated by the Earth. We incorporate the mass scale $\mu$ that has emerged in the previous Section \ref{sec:contraction} from the contraction of the $\kappa$-Poincar\'e algebra as deformation parameter of the Gallilei algebra, by means of first-order corrections to the spacetime symmetry transformation of the three-body system and to the Hamiltonian governing the interaction. We constrain the form of the interaction and the transformations by requiring that the potential is a relativistic invariant.

Our aim is to use this model in the next Section \ref{sec:EotvosDSR}, to quantify deviations from the universality of free fall by measuring the time of flight between two different masses, labeled A and B, falling towards the surface of the Earth, labeled E. The test particles' initial conditions are such that they fall from the same height and with zero initial velocity relatively to each other and to an observer that is initially located at the same position as the masses.
We assume that the test particles are affected by Planck-scale corrections while the Earth is treated as a classical Galilean particle. To our knowledge, this is only the second time that a model consisting of particles enjoying both deformed and undeformed symmetries is considered in the literature \cite{Amelino-Camelia:2019cjb}. 

The single-particle generators of particles with masses $m_A,m_B$ are denoted by $\{\vec P_A,\vec N_A,\vec R_A\}$ and $\{\vec P_B,\vec N_B,\vec R_B\}$, respectively, and we assume that the Poisson brackets for the single test-particle generators take the form \eqref{eq:algfirstorder}, whereas the kinetic energy for these particles is given by the on-shell relation displayed in \eqref{eq:kinfirstorder}. The Earth is treated as a classical Galilean particle of mass $m_E$ and its single-particle generators denoted by $\{\vec{P}_E,\vec{N}_E,\vec{R}_E\}$, close the Galilean algebra, in which the non-zero commutators read
\begin{equation}
\label{eq:Galileialg}
    \{R_{E}^i,P_{E}^j\}=\epsilon^{ijk}P_{E}^k \,, \qquad \{N_E^i,P_E^j\}=\delta^{ij}m_E \,, \qquad \{R_E^i,N_E^j\}=\epsilon^{ijk}N_E^k\, , \qquad \{R_E^i,R_E^j\}=\epsilon^{ijk}R_E^k \, .
\end{equation}
We denote the phase space coordinates of the test particles by $\vec x_A,\vec x_B,\vec p_A,\vec p_B$, respectively, and the phase space coordinates of the Earth by $\vec x_E,\vec p_E$. Their defining Poisson brackets are given by
\begin{equation}
    \{ x_Y^i,p_Z^j\}=\delta^{ij}\delta_{YZ}\, ,\qquad \qquad Y,Z=A,B,E, \quad i,j=1,2,3
\end{equation}
In terms of the phase space variables $x^i_I,p^j_J$ we can represent the symmetry generators. The generators of the deformed symmetry algebra in \eqref{eq:algfirstorder} can be written as\footnote{The time dependent part of the boost generator is found requiring that $\frac{dN_{I}^i}{dt}=\{N_{I}^i,H^{\rm free}_{I}\}$ where $H^{\rm free}_{I}=\frac{\vec p_{I}^2}{2m_{I}}(1+\epsilon_{I})$.}
\begin{equation}
\label{eq:symm_reps}
     P_{I}^i= p^i_{I}\,, \qquad R_{I}^i=\epsilon^{ijk}x_{I}^jp^k_{I} \qquad N_{I}^i=(1-\epsilon_{I})m_{I}x_{I}^i+p_{I}^i t\, ,
\end{equation}
with $I=A,B,$ whereas the symmetry generators for the Earth, satisfying \eqref{eq:Galileialg}, assume the standard Galilean expressions:
\begin{equation}
    P_E^i=p_E^i\,, \qquad R_{E}^i=\epsilon^{ijk}x_E^jp_E^k\,,\qquad N_E^i=m_Ex_E^i+p_E^it\,. 
\end{equation}

Inspired by the undeformed case, we aim to construct a Hamiltonian for gravitational interactions of particles $m_A,m_B$ with the Earth, which is invariant under the transformations generated by the total translation, total rotation and total boost of the system. Given that we are considering a system composed of two test-particles which enjoy deformed relativistic symmetries and a particle which enjoys undeformed Galilean symmetries, we make an ansatz for the expressions of the conserved charges of the total system at first order in the deformation parameter, satisfying the following requirements:
\begin{itemize}
\item The deformation terms are analytical and maintain dimensional consistency.
    \item Spatial isotropy is undeformed. 
    \item We do not allow for deformation terms involving quantities related solely to one particle, so we recover the definition of a single particle charge when the charges of the other particles are zero (\textit{e.g.}, we avoid terms of the form $\epsilon_I p_I$ in $p_{tot}$).
    \item Deformation terms are only proportional to $\epsilon_I=m_I/\mu$. If $\mu$ is supposed to be of the order of the Planck mass, corrections of the type $m_E/\mu$ would be unphysical, given that $m_E\gg \mu$, resulting in corrections of order $10^{32}$, which are in conflict with the observation that Newtonian gravity describes every-day phenomena on Earth scales.
\end{itemize}
Following these prescriptions, the most general parametrization of the composition laws of momentum, boost, angular momentum is 
\begin{equation}
\label{eq:totalgenApp}
    \begin{aligned}
    &  \ptot=\vec{p}_A+\vec{p}_B+\vec{p}_E+\epsilon_A(\alpha_1 \vec{p}_B+\alpha_2\vec{p}_E)+\epsilon_B(\alpha_3\vec{p}_A+\alpha_4\vec{p}_E) \, ,\\ &\Ntot=\vec{N}_A+\vec{N}_B+\vec{N}_E+\epsilon_A(\beta_1\vec{N}_B+\beta_2\vec{N}_E)+\epsilon_B(\beta_3\vec{N}_A+\beta_4\vec{N}_E) \, ,\\
    &  \Rtot=\vec{R}_A+\vec{R}_B+\vec{R}_E \, .
    \end{aligned}
\end{equation}
where $\alpha_n$ and $\beta_n$ ($n=1,2,3,4$) are real constants. In general, the composition of masses may be deformed as well, as is discussed for example in \cite{Ballesteros:1999ew}. Here, we omit the mass  composition because it does not influence the results of this paper.

In general, the momentum and boost addition laws are asymmetric in the particles $A$ and $B$, which is a common feature in general DSR models. However, an asymmetric addition law \emph{a priori} renders particles distinguishable even if they have equal properties (mass, spin, charge, \dots). To prevent this behaviour and for simplicity, in the main text of the paper we focus on DSR models with symmetric addition laws assuming that $\alpha_1=\alpha_3$, $\alpha_2=\alpha_4$, $\beta_1=\beta_3$, $\beta_2=\beta_4$. Then, the composition laws become
\begin{equation}
\label{eq:totalgen}
    \begin{aligned}
    &  \ptot=\vec{p}_A+\vec{p}_B+\vec{p}_E+\epsilon_A(\alpha_1 \vec{p}_B+\alpha_2\vec{p}_E)+\epsilon_B(\alpha_1\vec{p}_A+\alpha_2\vec{p}_E) \, ,\\
    &\Ntot=\vec{N}_A+\vec{N}_B+\vec{N}_E+\epsilon_A(\beta_1\vec{N}_B+\beta_2\vec{N}_E)+\epsilon_B(\beta_1\vec{N}_A+\beta_2\vec{N}_E) \, ,\\
    &  \Rtot=\vec{R}_A+\vec{R}_B+\vec{R}_E \, .
    \end{aligned}
\end{equation}
Nothing obstructs constructing a gravitational interaction potential and analyzing the universality of free fall in DSR theories with asymmetric composition laws. We present this more general case in Appendix \ref{app:MDR}, in order to not complicate our main insights  with additional parameters.

The model we construct shall satisfy the relativity principle. This enforces that conserved charges (as displayed in \eqref{eq:totalgen}) in one reference frame are conserved in all reference frames connected by relativistic transformations specified by \eqref{eq:algfirstorder}, \eqref{eq:Galileialg} and the total symmetry generators \eqref{eq:totalgen} \cite{Amelino-Camelia:2011gae,Amelino-Camelia:2013sba}. For spatial momentum conservation, we introduce $P_{\rm tot}^{i(f)}$  the initial (final) total momentum and label all other appearing momenta accordingly with a superscript $i$ or $f$. Then, $P_{\rm tot}^{i}=P_{\rm tot}^{f}$ implies
\begin{align}
\label{eq:momcons}
\vec{P}^i_{\rm tot}
&=\vec{p}_A^i+\vec{p}_B^i+\vec{p}_E^i+\epsilon_A(\alpha_1 \vec{p}_B^i+\alpha_2\vec{p}_E^i)+\epsilon_B(\alpha_1\vec{p}_A^i+\alpha_2\vec{p}_E^i)\nonumber\\
&=\vec{p}_A^f+\vec{p}_B^f+\vec{p}_E^f+\epsilon_A(\alpha_1 \vec{p}_B^f+\alpha_2\vec{p}_E^f)+\epsilon_B(\alpha_1\vec{p}_A^f+\alpha_2\vec{p}_E^f)
=\vec{P}^f_{\rm tot}\,.
\end{align}
To ensure compatibility with the notion of relativity introduced by the DSR transformations, it is necessary that the momentum-conservation equation \eqref{eq:momcons} is covariant under the action of the total generators written in \eqref{eq:totalgen}. For example, compatibility with an infinitesimal boost transformation of parameter $\vec{v}$ requires

\begin{equation}
\label{eq:totmomtransf}
   \vec {P}^{'i}_{\rm tot}=\vec{P}^i_{\rm tot}+ \{\vec{v}\cdot\vec{N}^i_{\rm tot},\vec{P}^i_{tot}\}=\vec{P}^f_{\rm tot}+ \{\vec{v}\cdot\vec{N}^f_{\rm tot},\vec{P}^f_{tot}\}=\vec P_{\rm tot}^{'f}.
\end{equation}

Using the relevant Poisson brackets in \eqref{eq:algfirstorder} and \eqref{eq:Galileialg}, it can be checked that this relativistic condition imposes no constraints on the parameters appearing in the charges. Analogous computations can also be done for the remaining generators  with equal outcome. Thus, our three particle model specified by the Poisson brackets \eqref{eq:algfirstorder} for the two test-particles, the Galilean symmetries \eqref{eq:Galileialg} for the Earth and the composition laws of the total conserved charges \eqref{eq:totalgen} is guaranteed to be relativistic.

The Hamiltonian governing the dynamics of the system can be written in the following way
\begin{equation}
    H=H_A^{\rm free}+H_B^{\rm free}+H_E^{\rm free}+V_A+V_B \, ,
\end{equation}
where the kinetic terms for the two test particles are given by \eqref{eq:defkinenergy} while the one for the Earth is the standard Galilean one: 
\begin{equation}\label{eq:KinTerms}
    H_{I}^{\rm free}=\frac{\vec{p}_{I}^2}{2m_{I}}(1+\epsilon_{I}) \, , \qquad H_E^{\rm free}=\frac{\vec{p}_E^2}{2m_E}\,.
\end{equation}
We could have chosen a different deformation of the dispersion relation compatible with the deformed symmetries. However, since this introduces no conceptually new ingredients, we relegate that analysis to Appendix~\ref{app:MDR}.

The potential terms $V_I$ characterize the interaction between particle $I$ and the Earth. As we are in the test-particle approximation, we neglect any interaction between particles $A$ and $B$ as well as the gravitational backreaction of the test particles on Earth. Requiring that the potential terms reduce to the standard Galilean invariant Newtonian interaction in the limit of vanishing deformation parameter $\epsilon_I$, we can write, at first order in $\epsilon_I$

\begin{equation}
   V_I=-\frac{G m_I m_E}{\abs{\vec{x}_I-\vec{x}_E}}(1+\epsilon_I f(\vec{x}_I,\vec{x}_E)) \,,
\end{equation}
where $f$ is a dimensionless function depending on the positions of the particles and the Earth.\footnote{In an asymmetric model the function $f$ can be distinct for the two particles. We discuss this possibility in Appendix \ref{app:MDR}.} The full interacting potential is then given by 
\begin{equation}
    V=\sum_{I=A,B}V_I.
\end{equation}

By requiring that the potential is invariant under the action of the total generators \eqref{eq:totalgen} at first order in $\epsilon_I$, namely that $\{\ptot,V\}=\{\Ntot,V\}=\{\Rtot,V\}=0$, we obtain constraints on the parameters, namely 
\begin{equation}
    \alpha_1=\alpha_2=\beta_1=\beta_2\equiv \alpha,\label{eq:ParameterIdentifications}
\end{equation}
where we summarized the kinematical deformation in the parameter $\alpha$ in the last equality. Furthermore, the form of $f$ is constrained to be
\begin{equation}
    f(\vec x_I,\vec x_E)=\frac{\alpha\, \vec{x}_I\cdot(\vec{x}_E-\vec{x}_I)}{\abs{\vec{x}_I-\vec{x}_E}^2}+h(\abs{\vec{x}_I-\vec{x}_E})\, ,
\end{equation}
where the dimensionless function $h$ can only depend on the respective Galilean distances $|\vec x_I-\vec x_E|$. The only scale available to construct the function $h$ has units of mass. Thus, by dimensional analysis we cannot balance any factor of $|\vec x_I-\vec x_E|,$ implying that the function $h$ has to be a dimensionless constant, which we continue to denote as $h$. In light of these constraints, the respective potentials can be rewritten as
\begin{equation}\label{eq:VI}
    V_I=-\frac{G m_I m_E}{\abs{\vec{x}_I-\vec{x}_E}}\qty(1+\epsilon_I\bigg( \frac{\alpha \, \vec{x}_I\cdot(\vec{x}_E-\vec{x}_I)}{\abs{\vec{x}_I-\vec{x}_E}^2}+h\bigg)) \, .
\end{equation}

The potential \eqref{eq:VI} is an example of a larger class of invariant interactions. Similarly to Galilean relativity, we can define quantities $d_I\equiv \abs{\vec{x}_I-(1-\epsilon_I\alpha_I))\vec{x}_E}$ that are invariant under translations, boosts and rotations.\footnote{Though as $\mu\to\infty$ the functions $d_I$ recover the Galilean distance, they are not distances (\eg $d_{I}(\vec{a},\vec{a})\neq0$) in the usual sense.} Thus, any invariant potential of particle $I$ interacting with the Earth has to be a function of $d_I$ only. As there is no constant with units of distance, which could balance powers of $d_I$, in form the modified potential has to depend on $d_I$ just as the ordinary potential depends on the distance. Therefore, in general the deformed Newtonian potential must take the shape 
\begin{equation}
    V_I\simeq-u(\epsilon_I)\frac{G m_I m_E}{d_I},\label{eq:PotetialDefDist}
\end{equation}
where we introduced the function $u$ satisfying $\lim_{\epsilon_I\to0}u(\epsilon_I)=1.$ In the parameterization of \eqref{eq:VI}, we have $u=1+\epsilon_I (h-\alpha)$.

To summarize, the total Hamiltonian for the three-body system now reads 
\begin{equation}
\label{eq:finHam}
    H= \sum_I\left[\frac{\vec{p}_I^2}{2m_I}(1+\epsilon_I)+V_I\right]+\frac{\vec{p}_E^2}{2m_E} \, .
\end{equation}
Given this Hamiltonian, we investigate the consequences for the universality of free fall in the following section.

\section{The tower of Pisa experiment in DSR\label{sec:EotvosDSR}}
  From the Hamiltonian \eqref{eq:finHam} the equations of motion for the two-test particles follow as 
\begin{equation}
\label{eq:acc_poisson}
    \ddot{\vec{x}}_I=\{\{\vec{x}_I,H\},H\}. 
\end{equation}
In the following, we  neglect the motion of the Earth given that its acceleration is going to be much smaller than the one of the test-particles, due to the fact that $m_A,m_B\ll \mu\ll m_E$.  In other words, we neglect backreaction as expected from test particles. To discuss the fate of free fall, we now focus on the following scenario. 

\begin{figure}[h!]
    \includegraphics[width=0.4\textwidth]{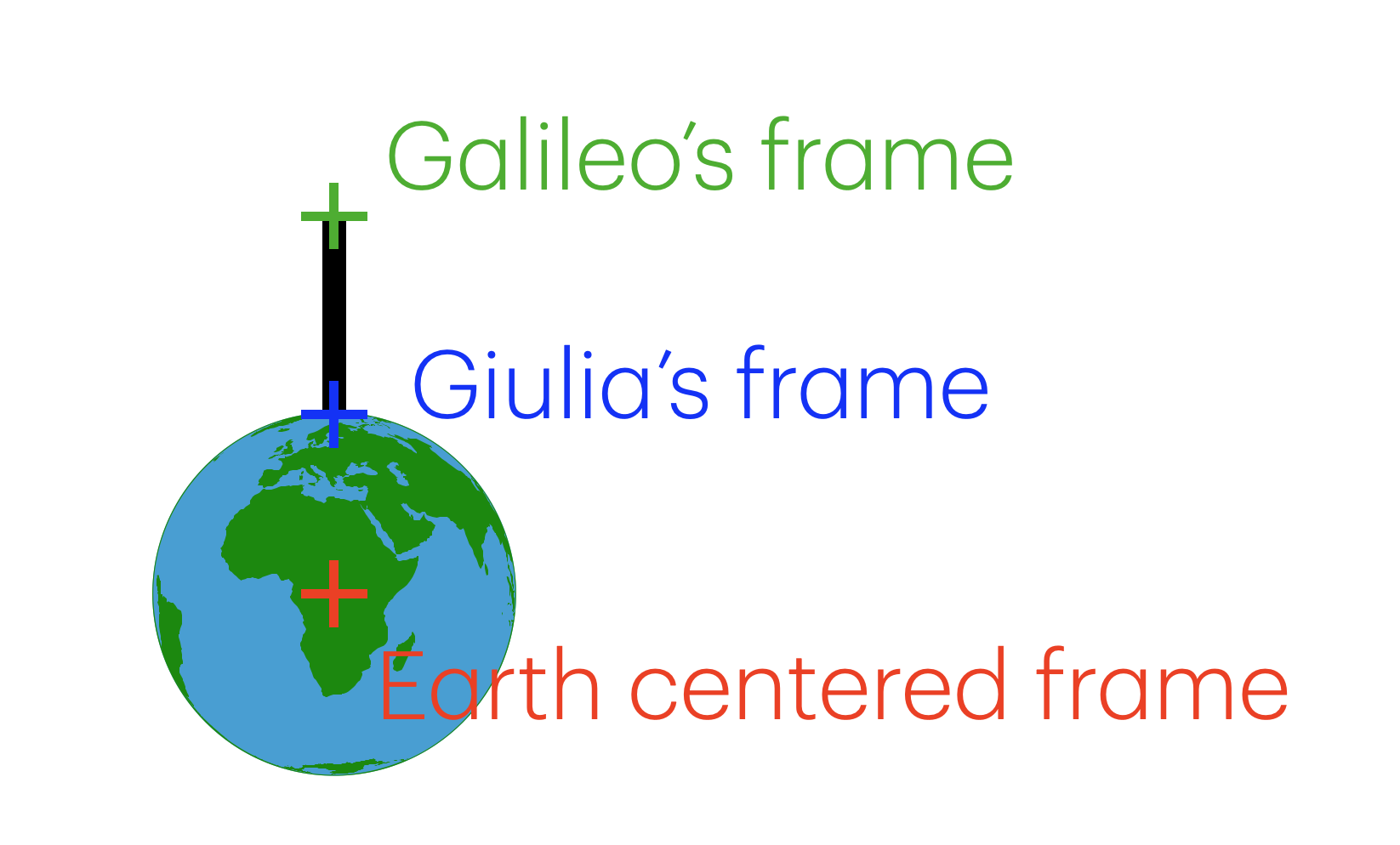}
    \caption{Sketch of the positions of the origins of the reference frames used in the main text.}
    \label{fig:frames}
\end{figure}

We consider two observers at relative rest, Galileo and Giulia, who want to perform the leaning-tower-of-Pisa experiment, see Figure \ref{fig:frames} for a schematic overview. Galileo is at the top of the tower holding the test particles, and lets them drop simultaneously with zero initial velocity towards Giulia, who detects them in her origin at the bottom of the tower. In this sense, Galileo is local to the release of the two particles, while Giulia is local to their detection. By measuring the time of arrival of the two objects, Giulia can quantify possible deviations from the universality of free fall. The frames are translated with respect to each other, which due to the deformed symmetries introduces relative-locality effects which are absent in Poincar\'e invariant theories.

As the origin of all three reference frames of Figure \ref{fig:frames} can be chosen to lie on the same axis, we effectively work in 1 dimension. Furthermore, thanks to the deformed translation invariance of the model, we conveniently choose to work in the Earth centered reference frame. We denote quantities in this reference frame with a superscript E. Under these assumptions, the expression for the potential terms \eqref{eq:VI} becomes:
\begin{equation}\label{eq:potentialE}
    V_{I}^E=-[1+\epsilon_{I}(h-\alpha)]\frac{Gm_{I}m_E}{x^E_{I}}.
\end{equation}
As a result, the equations of motion become
\begin{equation}
    \ddot x_{I}^E=-[1+\epsilon_{I}(h+1-\alpha)]\frac{Gm_E}{(x^E_{I})^2}\equiv -\bar g_{I}\left(\frac{\bar x_{I}^E}{x_{I}^E}\right)^2,\label{eq:EOMEarthFrame}
\end{equation}
where in the last equality we defined the initial acceleration $\bar g_I = \ddot x_I^E(0)$ depending on the initial position of the test particles in the Earth frame $\bar x_I^E\equiv x_I^E(0).$ The resulting equations of motion differ from the undeformed case just by the prefactor $[1+\epsilon_I(-h_I-1+\alpha_I)]$. Given the initial conditions $x_I^E(0)=\bar x_I^E$ and $\dot x_I^{E}(0)=0,$ the equations of motion \eqref{eq:EOMEarthFrame} are solved implicitly by solutions to the equation
\begin{equation}\label{eq:tI}
    t_{I}(x^E_I)=\frac{1}{\sqrt{2 \bar g_{I}}}\left[\bar x_I^E\arctan\sqrt{\frac{\bar x_{I}^E}{x_{I}^E}-1}+x_{I}^E\sqrt{\frac{\bar x_{I}^E}{ x_{I}^E}-1}\right].
\end{equation}
With the initial time normalized to zero, the time elapsed during the experiment is $t_{I}$.
We wish to evaluate the time of arrival at the bottom of the tower, in Giulia's reference frame, local to the particle detector. The particle trajectories in Giulia's reference frame can be obtained by a finite translation with parameter $\xi_G$ of the trajectories in the reference frame anchored to the center of the Earth. These transformations read
\begin{equation}
x_{I}^E(\bar t_{I})=\left.(x_{I}^G(\bar t_{I})-\xi_G\{p_{tot},x_{I}\})\right|_{x^G_I=0}=\xi_G(1+\epsilon_{J}\alpha)\,, \qquad I\neq J.\label{eq:frametrafo1}
\end{equation}
This relation implicitly defines a specific time $\bar t_I = t_I(x_{I}^E)|_{x_I^G=0}$, namely the time when Giulia detects the particle in the center of her rest frame. Thus, the time delay Giulia measures in her frame at the bottom of the tower is given by 
\begin{equation}\label{eq:td}
    \Delta t=\bar t_{A}-\bar t_B\,.
\end{equation}
At the same time, the initial positions for particles $A$ and $B$ in the Earth frame are related as
\begin{equation}\label{eq:frametrafo2}
    \bar x_{B}^E=\bar x_A^E\left[1+\alpha(\epsilon_A-\epsilon_B)\right].
\end{equation}
Using \eqref{eq:frametrafo1} and \eqref{eq:frametrafo2} in the time delay \eqref{eq:td} evaluated with \eqref{eq:tI}, the zeroth order vanishes, as expected from ordinary Newtonian gravity. Yet, corrections appear to first order. We obtain 
\begin{align}\label{eq:DeltaT}
    \Delta t=\frac{1+2\alpha+h}{2\sqrt{2g_0 x_0}}\left(\epsilon_A-\epsilon_B\right)\left[\xi_G\sqrt{\frac{x_0}{\xi_G}-1}+x_0\arctan\sqrt{\frac{x_0}{\xi_G}-1}\right]\,,
\end{align}
where we introduced the 0th-order short-hand abbreviations $x_0\equiv \bar x_{A}^E=\bar x_{B}^E+\mathcal{O}(\epsilon_I)$ for the initial position of the particles $I$ (the top of the tower), and $g_0\equiv Gm_E/x_0^2$ as their classical initial acceleration. Note that, as $\Delta t$ is a first-order expression, $\xi_G$ in \eqref{eq:DeltaT}  approximates  Giulia's classical position. 
 
We denote the classical free-fall time as $T.$ For short runtimes, \ie $|g_0T^2/2x_0|\ll1$, Giulia's position equals $\xi_G\simeq x_0-g_0T^2/2$. Then, we obtain the approximation
\begin{equation}
    \label{eq:td2}
    \Delta t\simeq\frac{T}{2}(1+2\alpha+h)(\epsilon_A-\epsilon_B).
\end{equation}
For general choices of the deformation parameters $\alpha_I,h_I$, the time delay \eqref{eq:td} is non-zero, so when particle $A$ is in the origin of the observer performing the detection, particle $B$ is not, and viceversa, signaling a violation of the universality of free fall.

However, a closer look at the free coefficients involved shows that we can preserve the universality of free fall in the deformed-relativity scenario if and only if 
\begin{equation}
\label{eq:constrainteffectiveEotvos0}
    h=-(1+2\alpha)\,.
\end{equation}
Thus, the universality of free fall requires the following form of the potential
\begin{equation}\label{eq:DSRnewton}
    V_I=-\frac{G m_I m_E}{\abs{\vec{x}_I-\vec{x}_E}}\qty(1-\epsilon_I\bigg(1+2\alpha- \frac{\alpha \, \vec{x}_I\cdot(\vec{x}_E-\vec{x}_I)}{\abs{\vec{x}_I-\vec{x}_E}^2}\bigg)) \, .
\end{equation}\vspace{11pt}

To summarize, we have found that for generic DSR models the universality of free fall is violated. However, one can construct special DSR models with a modified Newtonian potential, such that the universality of free fall is preserved.\vspace{11pt}

As a last point of our study, we investigate the behavior of the time delay under boost transformations. 
Up to now, we have considered the case in which Galileo and Giulia are at relative rest. We now want to investigate what happens when Giulia moves towards Galileo (or the converse). In the undeformed case, the particles emitted by Galileo evidently still arrive at Giulia's detector simultaneously. 

For definiteness, let us focus on the scenario in which Galileo is at relative rest with respect to the Earth, and Giulia is boosted with respect to both. Again, we work in the Earth reference frame, in which the expression for the time of flight is still given by \eqref{eq:td}.
In order to evaluate the time of flight in Giulia's reference frame, we act with both a finite translation and a finite boost (with parameter $v$) on the test particles' worldlines:
\begin{equation}\label{eq:genframetrafo2}
x_{I}^E(\bar t_{I})=\left.(x_{I}^G(\bar t_{I})-\xi_G\{p_{tot},x_{I}\}-v\{N_{tot},x_{I}\})\right|_{x^G_I=0}=(\xi_G+vt_I)(1+\epsilon_{J}\alpha_{J})\,,\quad I\neq J.
\end{equation}
Again, the above relation defines a time $\bar t_I=t_I(x_I^E)|_{x_G^I=0}$, where Giulia detects particle $I$ in the origin of her rest frame. The relation between the initial positions for particles $A,B$ in the Earth frame is still given by \eqref{eq:frametrafo2}, since it does not depend on the state of Giulia's reference frame. The final result, expanded to first order in $\epsilon_I$, is given by
\begin{equation}
\begin{aligned}\label{eq:DeltaTBoost}
    \Delta t=&\frac{(1+2\alpha+h)}{2(\sqrt{2g_0 x_0}+\frac{v}{\sqrt{\frac{x_0}{\xi_G+v T}-1}})}\left(\epsilon_A-\epsilon_B\right)\\
    &\times \left[(\xi_G+vT)\sqrt{\frac{x_0}{\xi_G+v T}-1}+x_0\arctan\sqrt{\frac{x_0}{\xi_G+v T}-1}\right].
    \end{aligned}
\end{equation}
We see that even in the case where Galileo and Giulia are boosted with respect to each other, the condition ensuring a vanishing time delay is the same as the one obtained when our observers are at relative rest. This implies that if the modification to the Newtonian potential ($h_I$) is chosen such that  \eqref{eq:constrainteffectiveEotvos0} is satisfied, universality of free fall holds in every frame of reference. As a consistency test, we have checked that this remains true in the  complementary case that Giulia is at rest with respect to the Earth while Galileo is moving. 

Gravity is absent in the limit $g_0\to 0$. In that limit, Giulia's position equals $\lim_{g_0\to0}(\xi_G+vT) = x_0$ and we see that \eqref{eq:DeltaTBoost} implies 
\begin{equation}
    \lim_{g_0\to 0}\Delta t=0\,.\label{eq:NoGravityNoTimeDelay}
\end{equation}
Notably, \eqref{eq:NoGravityNoTimeDelay} is not in conflict with \eqref{eq:DeltaT} which becomes trivial when $g_0\to0$ -- without initial velocity or acceleration, the test masses do not move.

Thus, particles with distinct masses but equal initial boost arrive without time delay. This is an immediate implication of the parameter constraints \eqref{eq:ParameterIdentifications}. In other words, demanding consistency with the existence of a potential in our model already requires universality of free fall in the absence of gravity and precludes observable time delays altogether (in the absence of other interactions), which is a remarkable feature for a deformed relativity model. With this observation we have found a deformed Galilean relativity model which possesses a modified dispersion relation, and realized the WEP. 

In short, universality of free fall in all reference frames can be guaranteed if and only if $h=-1-2\alpha.$ Thus, it fixes all free parameters in the Hamiltonian -- only the model-characteristic, kinematical parameter $\alpha$ remains. Furthermore, with the identifications given in \eqref{eq:ParameterIdentifications}, the time delay always vanishes in the absence of gravity. Thus, in our model free fall is truly universal.

\section{Summary, discussion and outlook}
\label{sec:discussion}
The WEP was and is one of the key guiding principles for the construction of theories of gravity. Motivated by the question if it carries on to the regime of quantum fields and quantum gravity we have studied the universality of free fall in the Galilean and Newtonian limit of DSR theories.

The main new results of our investigation are:
\begin{itemize}
    \item The construction of a new family of DSR models (in the Galilean/Newtonian regime) in which two test particles subject to the deformed symmetry algebra, here the deformed Galilei algebra, interact with a background gravitational field of a classical source, here the Earth. The Hamiltonian defining this theory is given by \eqref{eq:finHam} and includes DSR modifications of the Newtonian potential. 
    \item A key prediction of this model is that the universality of free fall does in general not hold. Two particles, which are simultaneously set into free fall in their rest frame do not arrive simultaneously at an observer at a distance who detects the particles. The test particles acquire a time delay, see \eqref{eq:DeltaT} and \eqref{eq:DeltaTBoost}.
    \item  Remarkably, we find that there exist particular DSR models (in the Galilean/Newtonian regime), for which the free parameters can be chosen such that the universality of free fall holds \eqref{eq:constrainteffectiveEotvos0}. These DSR models necessarily predict a modification of the Newtonian potential \eqref{eq:DSRnewton}.
    \item The requirement of being able to construct an invariant potential constrains available models as found in equation~\eqref{eq:ParameterIdentifications}. 
    In our model this constraint suffices to ensure universality of free fall in the absence of gravity, as seen from equation \eqref{eq:NoGravityNoTimeDelay}. 
    Thus, adding interactions forces us to use a Galilean  DSR model with MDR which does not lead to observationally relevant time delays even for free particles sharing the same initial conditions. Whether this finding generalizes to other models, will be the subject of future work.
\end{itemize}
Experiments testing the universality of free fall \cite{Will:2014kxa, MICROSCOPE:2022doy} usually measure the E\"otv\'os factor, quantifying the difference in gravitational acceleration of two test bodies. Let us investigate how this indicator performs in our DSR model. Defining the E\"otv\'os factor in terms of the coordinate accelerations of the two masses evaluated in the same point through
\begin{align}
\label{eq:eotvos_coordinate}
    \eta = 2\left|\frac{\ddot x_A - \ddot x_B}{\ddot x_A+\ddot x_B}\right| \,,
\end{align}
we obtain, to first order in $\epsilon_I$ in the Earth frame\footnote{It can be checked explicitly using relation \eqref{eq:acc_poisson} for the acceleration that the E\"otv\'os factor is invariant under the deformed symmetry transformations, \textit{i.e.}, $\{p_{\rm tot},\eta\}=\{N_{\rm tot},\eta\}=\{R_{\rm tot},\eta\}=0$, thus yielding the same result in any reference frame.}
\begin{align}
    \eta &= 2\left|\frac{x_A^E-x^E_B}{x_A^E+x^E_B}- \frac{2 x_A^E  x_B^E (h+1-\alpha)(\epsilon_A-\epsilon_B)}{(x_A^E+x^E_B)^2}\right|\,.
\end{align}
Using the fact that $x_B^E=x_A^E(1+\epsilon_A\alpha-\epsilon_B\alpha)$, we find at first order
\begin{equation}
    \eta=|(1+h+\alpha)(\epsilon_B-\epsilon_A)| \, ,\label{eq:CoordEotvos}
\end{equation}
which vanishes for $h=-(1+\alpha)$ and not for the universality-of-free-fall condition $h=-(1+2\alpha)$ obtained in Section~\ref{sec:EotvosDSR}. Of course, in the limit $\epsilon_I\rightarrow 0$, both the time delay analysis of Section \ref{sec:EotvosDSR} and the E\"otv\'os-factor analysis are equivalent in their conclusions that free fall is universal. The discrepancy between the two approaches in the DSR case stems from the fact that the E\"otv\'os factor analysis fails to take into account the relative-locality effects generated by the deformed symmetry transformations needed to describe the detection events local to Giulia. Not incidentally, if and only if  $\alpha=0$, corresponding to the case of trivial composition of symmetry generators (see Eqs. \eqref{eq:totalgen}, \eqref{eq:ParameterIdentifications}), the two analyses lead to the same condition $h=-1$ for universality of free fall to hold.

Physically speaking, a model in which $h=-(1+2\alpha)$ predicts no time delay in the leaning tower of Pisa experiment, but a non-vanishing E\"otv\'os factor. Hence, the latter proves to be an unsuitable estimator of deviations from the universality of free fall in the context of DSR models.

Besides the coordinate E\"otv\'os factor, for Earth-based experiments we can define an effective E\"otv\'os factor that corresponds to the interpretation of the experiment in an undeformed Galilean-relativity context. Given that for terrestrial scales the distance $D$ traveled by test particle $I$ is much smaller than the radius of the Earth $R_E$ (typically $D/R_E\sim 10^{-5}$), in the underformed case we can infer the acceleration of the test particle, $a_I$, in terms of $D$ and its travel time $t_I$ by means of
\begin{align}
    a_I = \frac{2D}{t^2_I}+\mathcal{O}\left(\frac{D}{R_E}\right)\ ,
\end{align}
where we assumed no initial velocities for the test particles.\footnote{For definiteness we focus on the case where the observers are at relative rest. It can be checked that when they are boosted relative to each other, the effective E\"otv\'os factor remains the same at the order we are considering.}
Based on the time delay $\Delta t = t_A-t_B$ Giulia observes, one can construct the E\"otv\'os factor from the effective accelerations
\begin{align}
    \eta_{\rm eff} = 2\frac{|a_A-a_B|}{|a_A+a_B|} 
    =  2\left|\frac{\frac{1}{t_A} - \frac{1}{t_B}}{\frac{1}{t_A} + \frac{1}{t_B}}\right|  
    = 2\frac{|\Delta t|}{|2t_B + \Delta t|}\,.
\end{align}
This effective E\"otv\'os factor is directly related to the validity of the universality of free fall and should be compared with measurements instead of \eqref{eq:CoordEotvos}. In our approximation, we can expand the effective E\"otv\'os factor as
\begin{equation}
    \eta_{\rm eff}=\frac{1}{2}|(1+2\alpha+h)(\epsilon_A-\epsilon_B)|+\mathcal{O}\left(\frac{D}{R_E}\right) \, .
\end{equation}
This directly relates the effective E\"otv\'os factor with the model parameters.

Hence, constraints on the E\"otv\'os factor yield constraints on the model parameters. For the distinguished DSR models, for which the coefficients are chosen such that \eqref{eq:constrainteffectiveEotvos0} holds and there is no time delay, the effective E\"otv\'os factor vanishes. However, since in this case a modified Newtonian potential is unavoidable, we plan to investigate its influence on further observables in the future. 

In general, our construction of the deformed Galilean gravitational interaction offers many avenues for phenomenological testing for the different parameters appearing. Depending on the choice of parameters a time delay can be avoided and the universality of free fall can be implemented, however modifications of Newtons potential are always present as long as the deformation is non-trivial.

Particularly interesting are experiments which measure the value of Newtons constant \cite{Muller:2007zzb,Das:2024xgf} or testing the equivalence of inertial and gravitational mass \cite{MICROSCOPE:2022doy,Singh:2022wyp}, as these aspects change with respect to the undeformed Galilean theory when studying consequences of the deformed Hamiltonian \eqref{eq:finHam}. 

Our results serves as proof of principle  that the universality of free fall severely constrains DSR  models. An important next step is to extend the construction of interacting DSR particles with a background gravitational field to the fully relativistic regime, to study the equivalence principle beyond Newtonian quantum gravity phenomenology. Besides, given that we have found the generalization of the Newtonian potential to $\kappa$-deformed symmetries, it would be interesting to find a Galilean (possibly non-commutative) field theory of gravity, based on a generalization of Poisson's equation, which recovers our results. Furthermore, this generalized Poisson's equation may be derivable from a generalization of weak-field general relativity.

Beyond gravity, our construction paves the way to do a similar analysis for further interactions like electromagnetism. In particular, it would be interesting to see whether the presence of interactions requires vanishing time delays for a larger class of models as discussed at the end of Section \ref{sec:EotvosDSR}.

\section*{Acknowledgements}

The authors would like to acknowledge the contribution of the COST Action CA23130 (“Bridging high and low energies in search of quantum gravity (BridgeQG)”). G.F.'s work on this project was supported by ``The Foundation Blanceflor". CP acknowledges support by the excellence cluster QuantumFrontiers of the German Research Foundation (Deutsche Forschungsgemeinschaft, DFG) under Germany's Excellence Strategy -- EXC-2123 QuantumFrontiers -- 390837967 and was funded by the Deutsche Forschungsgemeinschaft (DFG, German Research Foundation) - Project Number 420243324

\appendix

\section{More general parameterization\label{app:MDR}}
In the main text we have considered a model with three free parameters, two governing the composition of boosts and momenta, respectively, and one characterizing the deformation of the gravitational potential. In this appendix, we extend the number of free parameters to demonstrate that the findings of the main text are robust. 

There are two ways in which we can increase the parameter space of our model. First, we can generalize the dispersion relation, \ie the kinetic term of the Hamiltonian of the $\kappa$-deformed particles. This yields one possible additional parameter per particle. Second, composition laws based on the the $\kappa$-Poincar\'e Hopf algebra are often not symmetric. Allowing for asymmetry as in \eqref{eq:totalgenApp} doubles the number of free parameters in the composition laws, but also in the dispersion relations and the potential.

We start with the dispersion relation. In the main text, we have chosen the kinetic energy for the deformed particles to equal \eqref{eq:defkinenergy}. This is a consequence of the Wigner-\.In\"on\"u contraction of the Casimir element \eqref{eq:Casimir element}. Yet any function of a Casimir is still a Casimir element for a symmetry algebra. Working at first order in $1/\kappa$, where the Casimir element \eqref{eq:Casimir element} assumes the expression $C=P_0^2-\vec{P}^2c^2(1+\frac{P_0}{\kappa})$, we could have equivalently chosen the particle-specific Casimir 
\begin{equation}
\label{eq:newcasimir}
    C_I'\simeq C+2(\gamma_I-1) \frac{C^{3/2}}{\kappa} \, ,
\end{equation}
where $\gamma_I$ are dimensionless parameters. At first order in the deformation, \eqref{eq:newcasimir} is the most general Casimir redefinition possible. Applying the Wigner-\.In\"on\"u contraction to the Casimir \eqref{eq:newcasimir}, we find the free-particle Hamiltonian
\begin{equation}
        H_I^0=\frac{\vec{p}_I^2}{2m_I}(1+\epsilon_I\gamma_I) \,.
\end{equation}

Next, we allow for the possibility that the particles are not treated symmetrically. To be consistent with the existence of an invariant potential, the additional free parameters are constrained. Following the logic of Section \ref{sec:DSRModel}, the constraints \eqref{eq:ParameterIdentifications} in the more general parameterization become
\begin{equation}
    \alpha_1=\alpha_2=\beta_1+1-\gamma_A=\beta_2+1-\gamma_A\equiv \alpha_A \, ,\qquad \alpha_3=\alpha_4=\beta_3+1-\gamma_B=\beta_4+1-\gamma_B\equiv \alpha_B \,,
\end{equation}
where the respective last equalities define the summarizing parameters $\alpha_A$ and $\alpha_B.$ Furthermore, the representation of the boost of the test particles \eqref{eq:symm_reps} in this case reads
\begin{equation} N_{I}^i=(1-\epsilon_{I})m_{I}x_{I}^i+(1+\epsilon_{I}(1-\gamma_I))p_{I}^i t\, .
\end{equation}
As the composition law is asymmetric, so can be the potential. We define it as
\begin{equation}\label{eq:VIApp}
    V_I=-\frac{G m_I m_E}{\abs{\vec{x}_I-\vec{x}_E}}\qty(1+\epsilon_I\bigg( \frac{\alpha_I \, \vec{x}_I\cdot(\vec{x}_E-\vec{x}_I)}{\abs{\vec{x}_I-\vec{x}_E}^2}+h_I\bigg)),
\end{equation}
where $h_I$ now define independent parameters for each $\kappa$-deformed particle.

Following the steps mapped out in Section \ref{sec:EotvosDSR}, we obtain the time delay in the non-boosted case:
\begin{align}
    \Delta t=\frac{1}{2\sqrt{2g_0 x_0}}\left[\epsilon_A(\gamma_A+2\alpha_A+h_A)-\epsilon_B(\gamma_B+2\alpha_B+h_B)\right]\left[\xi_G\sqrt{\frac{x_0}{\xi_G}-1}+x_0\arctan\sqrt{\frac{x_0}{\xi_G}-1}\right],
\end{align}
and in the boosted case: 
\begin{equation}
\begin{aligned}
   \Delta t=&\frac{1}{2(\sqrt{2g_0 x_0}+\frac{v}{\sqrt{\frac{x_0}{\xi_G+v T}-1}})}\left[\epsilon_A(\gamma_A+2\alpha_A+h_A)-\epsilon_B(\gamma_B+2\alpha_B+h_B)\right]\\
    &\times \left[(\xi_G+vT)\sqrt{\frac{x_0}{\xi_G+v T}-1}+x_0\arctan\sqrt{\frac{x_0}{\xi_G+v T}-1}\right].
    \end{aligned}
\end{equation}
The time delay vanishes when $h_I=-2\alpha_I-\gamma$ and reduces to \eqref{eq:DeltaT} and \eqref{eq:DeltaTBoost} in the limit $\gamma_I\rightarrow 1,$ $(\alpha_A,h_A),(\alpha_B,h_B)\to\alpha,h$  in the non-boosted and boosted case, respectively. 
We find that the additional parameters do not lead to qualitative changes of our main results.

\providecommand{\href}[2]{#2}\begingroup\raggedright\endgroup


\begin{thebibliography}{10}
	
	\bibitem{Addazi:2021xuf}
	A.~Addazi {\em et~al.}, ``{Quantum gravity phenomenology at the dawn of the
		multi-messenger era{\textemdash}A review},'' {\em Prog. Part. Nucl. Phys.}
	{\bfseries 125} (2022) 103948, \href{http://arxiv.org/abs/2111.05659}{
		arXiv:2111.05659 [hep-ph]}.
	
	\bibitem{Amelino-Camelia:2000stu}
	G.~Amelino-Camelia, ``{Relativity in space-times with short distance structure
		governed by an observer independent (Planckian) length scale},'' {\em Int. J.
		Mod. Phys. D} {\bfseries 11} (2002) 35--60,
	\href{http://arxiv.org/abs/gr-qc/0012051}{ arXiv:gr-qc/0012051}.
	
	\bibitem{Amelino-Camelia:2002cqb}
	G.~Amelino-Camelia, ``{Doubly special relativity},'' {\em Nature} {\bfseries
		418} (2002) 34--35, \href{http://arxiv.org/abs/gr-qc/0207049}{
		arXiv:gr-qc/0207049}.
	
	\bibitem{Magueijo:2001cr}
	J.~Magueijo and L.~Smolin, ``{Lorentz invariance with an invariant energy
		scale},'' {\em Phys. Rev. Lett.} {\bfseries 88} (2002) 190403,
	\href{http://arxiv.org/abs/hep-th/0112090}{ arXiv:hep-th/0112090}.
	
	\bibitem{Kowalski-Glikman:2002iba}
	J.~Kowalski-Glikman and S.~Nowak, ``{Doubly special relativity theories as
		different bases of kappa Poincare algebra},'' {\em Phys. Lett. B} {\bfseries
		539} (2002) 126--132, \href{http://arxiv.org/abs/hep-th/0203040}{
		arXiv:hep-th/0203040}.
	
	\bibitem{Amelino-Camelia:1999jfz}
	G.~Amelino-Camelia and S.~Majid, ``{Waves on noncommutative space-time and
		gamma-ray bursts},'' {\em Int. J. Mod. Phys. A} {\bfseries 15} (2000)
	4301--4324, \href{http://arxiv.org/abs/hep-th/9907110}{
		arXiv:hep-th/9907110}.
	
	\bibitem{Borowiec:2010yw}
	A.~Borowiec and A.~Pachol, ``{$\kappa$-Minkowski spacetimes and DSR algebras:
		Fresh look and old problems},'' {\em SIGMA} {\bfseries 6} (2010) 086,
	\href{http://arxiv.org/abs/1005.4429}{ arXiv:1005.4429 [math-ph]}.
	
	\bibitem{Matschull:1997du}
	H.-J. Matschull and M.~Welling, ``{Quantum mechanics of a point particle in
		(2+1)-dimensional gravity},'' {\em Class. Quant. Grav.} {\bfseries 15} (1998)
	2981--3030, \href{http://arxiv.org/abs/gr-qc/9708054}{ arXiv:gr-qc/9708054}.
	
	\bibitem{Freidel:2003sp}
	L.~Freidel, J.~Kowalski-Glikman, and L.~Smolin, ``{2+1 gravity and doubly
		special relativity},'' {\em Phys. Rev. D} {\bfseries 69} (2004) 044001,
	\href{http://arxiv.org/abs/hep-th/0307085}{ arXiv:hep-th/0307085}.
	
	\bibitem{Freidel:2005me}
	L.~Freidel and E.~R. Livine, ``{3D Quantum Gravity and Effective Noncommutative
		Quantum Field Theory},'' {\em Phys. Rev. Lett.} {\bfseries 96} (2006) 221301,
	\href{http://arxiv.org/abs/hep-th/0512113}{ arXiv:hep-th/0512113}.
	
	\bibitem{Freidel:2005bb}
	L.~Freidel and E.~R. Livine, ``{Ponzano-Regge model revisited III: Feynman
		diagrams and effective field theory},'' {\em Class. Quant. Grav.} {\bfseries
		23} (2006) 2021--2062, \href{http://arxiv.org/abs/hep-th/0502106}{
		arXiv:hep-th/0502106}.
	
	\bibitem{Cianfrani:2016ogm}
	F.~Cianfrani, J.~Kowalski-Glikman, D.~Pranzetti, and G.~Rosati, ``{Symmetries
		of quantum spacetime in three dimensions},'' {\em Phys. Rev. D} {\bfseries
		94} no.~8, (2016) 084044, \href{http://arxiv.org/abs/1606.03085}{
		arXiv:1606.03085 [hep-th]}.
	
	\bibitem{Bojowald:2012ux}
	M.~Bojowald and G.~M. Paily, ``{Deformed General Relativity},'' {\em Phys. Rev.
		D} {\bfseries 87} no.~4, (2013) 044044,
	\href{http://arxiv.org/abs/1212.4773}{ arXiv:1212.4773 [gr-qc]}.
	
	\bibitem{Amelino-Camelia:2016gfx}
	G.~Amelino-Camelia, {\em et~al.}, ``{Spacetime-noncommutativity regime of Loop
		Quantum Gravity},'' {\em Phys. Rev. D} {\bfseries 95} no.~2, (2017) 024028,
	\href{http://arxiv.org/abs/1605.00497}{ arXiv:1605.00497 [gr-qc]}.
	
	\bibitem{Veneziano:1986zf}
	G.~Veneziano, ``{A Stringy Nature Needs Just Two Constants},'' {\em EPL}
	{\bfseries 2} (1986) 199.
	
	\bibitem{Kostelecky:1988zi}
	V.~A. Kostelecky and S.~Samuel, ``{Spontaneous Breaking of Lorentz Symmetry in
		String Theory},'' {\em Phys. Rev. D} {\bfseries 39} (1989) 683.
	
	\bibitem{Yoneya:2000bt}
	T.~Yoneya, ``{String theory and space-time uncertainty principle},'' {\em Prog.
		Theor. Phys.} {\bfseries 103} (2000) 1081--1125,
	\href{http://arxiv.org/abs/hep-th/0004074}{ arXiv:hep-th/0004074}.
	
	\bibitem{Amelino-Camelia:1999iec}
	G.~Amelino-Camelia, J.~Lukierski, and A.~Nowicki, ``{Distance measurement and
		kappa deformed propagation of light and heavy probes},'' {\em Int. J. Mod.
		Phys. A} {\bfseries 14} (1999) 4575--4588,
	\href{http://arxiv.org/abs/gr-qc/9903066}{ arXiv:gr-qc/9903066}.
	
	\bibitem{Barcaroli:2015xda}
	L.~Barcaroli, {\em et~al.}, ``{Hamilton geometry: Phase space geometry from
		modified dispersion relations},'' {\em Phys. Rev. D} {\bfseries 92} no.~8,
	(2015) 084053, \href{http://arxiv.org/abs/1507.00922}{ arXiv:1507.00922
		[gr-qc]}.
	
	\bibitem{Barcaroli:2017gvg}
	L.~Barcaroli, {\em et~al.}, ``{Curved spacetimes with local
		$\kappa$-Poincar{\'e} dispersion relation},'' {\em Phys. Rev. D} {\bfseries
		96} no.~8, (2017) 084010, \href{http://arxiv.org/abs/1703.02058}{
		arXiv:1703.02058 [gr-qc]}.
	
	\bibitem{Pfeifer:2021tas}
	C.~Pfeifer and J.~J. Relancio, ``{Deformed relativistic kinematics on curved
		spacetime: a geometric approach},'' {\em Eur. Phys. J. C} {\bfseries 82}
	no.~2, (2022) 150, \href{http://arxiv.org/abs/2103.16626}{ arXiv:2103.16626
		[gr-qc]}.
	
	\bibitem{Kowalski-Glikman:2002eyl}
	J.~Kowalski-Glikman and S.~Nowak, ``{Noncommutative space-time of doubly
		special relativity theories},'' {\em Int. J. Mod. Phys. D} {\bfseries 12}
	(2003) 299--316, \href{http://arxiv.org/abs/hep-th/0204245}{
		arXiv:hep-th/0204245}.
	
	\bibitem{Colladay:1998fq}
	D.~Colladay and V.~A. Kostelecky, ``{Lorentz violating extension of the
		standard model},'' {\em Phys. Rev. D} {\bfseries 58} (1998) 116002,
	\href{http://arxiv.org/abs/hep-ph/9809521}{ arXiv:hep-ph/9809521}.
	
	\bibitem{Kostelecky:2011qz}
	A.~Kostelecky, ``{Riemann-Finsler geometry and Lorentz-violating kinematics},''
	{\em Phys. Lett. B} {\bfseries 701} (2011) 137--143,
	\href{http://arxiv.org/abs/1104.5488}{ arXiv:1104.5488 [hep-th]}.
	
	\bibitem{Amelino-Camelia:2011lvm}
	G.~Amelino-Camelia, L.~Freidel, J.~Kowalski-Glikman, and L.~Smolin, ``{The
		principle of relative locality},'' {\em Phys. Rev. D} {\bfseries 84} (2011)
	084010, \href{http://arxiv.org/abs/1101.0931}{ arXiv:1101.0931 [hep-th]}.
	
	\bibitem{Amelino-Camelia:2011hjg}
	G.~Amelino-Camelia, L.~Freidel, J.~Kowalski-Glikman, and L.~Smolin, ``{Relative
		locality: A deepening of the relativity principle},'' {\em Gen. Rel. Grav.}
	{\bfseries 43} (2011) 2547--2553, \href{http://arxiv.org/abs/1106.0313}{
		arXiv:1106.0313 [hep-th]}.
	
	\bibitem{Gubitosi:2011hgc}
	G.~Gubitosi and F.~Mercati, ``{Relative Locality in $\kappa$-Poincar{\'e}},''
	{\em Class. Quant. Grav.} {\bfseries 30} (2013) 145002,
	\href{http://arxiv.org/abs/1106.5710}{ arXiv:1106.5710 [gr-qc]}.
	
	\bibitem{Amelino-Camelia:2011uwb}
	G.~Amelino-Camelia, {\em et~al.}, ``{Relative-locality distant observers and
		the phenomenology of momentum-space geometry},'' {\em Class. Quant. Grav.}
	{\bfseries 29} (2012) 075007, \href{http://arxiv.org/abs/1107.1724}{
		arXiv:1107.1724 [hep-th]}.
	
	\bibitem{Brunkhorst:2017gcz}
	L.~K. Brunkhorst, {\em {Deformation and contraction of symmetries in special relativity}}, 11, 2017.
	\newblock PhD thesis, U. Bremen (main).
	
	\bibitem{Majid:1994cy}
	S.~Majid and H.~Ruegg, ``{Bicrossproduct structure of kappa Poincare group and
		noncommutative geometry},'' {\em Phys. Lett. B} {\bfseries 334} (1994)
	348--354, \href{http://arxiv.org/abs/hep-th/9405107}{ arXiv:hep-th/9405107}.
	
	\bibitem{Amelino-Camelia:2014gga}
	G.~Amelino-Camelia, ``{Planck-scale soccer-ball problem: a case of mistaken
		identity},'' {\em Entropy} {\bfseries 19} no.~8, (2017) 400,
	\href{http://arxiv.org/abs/1407.7891}{ arXiv:1407.7891 [gr-qc]}.
	
	\bibitem{Bosso:2023nst}
	P.~Bosso, G.~Fabiano, D.~Frattulillo, and F.~Wagner, ``{Fate of Galilean
		relativity in minimal-length theories},'' {\em Phys. Rev. D} {\bfseries 109}
	no.~4, (2024) 046016, \href{http://arxiv.org/abs/2307.12109}{
		arXiv:2307.12109 [gr-qc]}.
	
	\bibitem{Amelino-Camelia:2023rkg}
	G.~Amelino-Camelia, G.~Fabiano, and D.~Frattulillo, ``{Total Momentum and Other
		Noether Charges for Particles Interacting in a Quantum Spacetime},'' {\em
		Symmetry} {\bfseries 17} no.~2, (2025) 227,
	\href{http://arxiv.org/abs/2302.08569}{ arXiv:2302.08569 [hep-th]}.
	
	\bibitem{Hohmann:2024lys}
	M.~Hohmann, C.~Pfeifer, and F.~Wagner, ``{Weak equivalence principle and
		nonrelativistic limit of general dispersion relations},'' {\em Phys. Rev. D}
	{\bfseries 110} no.~10, (2024) 104030,
	\href{http://arxiv.org/abs/2404.18811}{ arXiv:2404.18811 [gr-qc]}.
	
	\bibitem{Arzano:2022nlo}
	M.~Arzano, V.~D'Esposito, and G.~Gubitosi, ``{Fundamental decoherence from
		quantum spacetime},'' {\em Commun. Phys.} {\bfseries 6} no.~1, (2023) 242,
	\href{http://arxiv.org/abs/2208.14119}{ arXiv:2208.14119 [gr-qc]}.
	
	\bibitem{Ballesteros:2020uxp}
	A.~Ballesteros, G.~Gubitosi, I.~Gutierrez-Sagredo, and F.~J. Herranz, ``{The
		$\kappa$-Newtonian and $\kappa$-Carrollian algebras and their noncommutative
		spacetimes},'' {\em Phys. Lett. B} {\bfseries 805} (2020) 135461,
	\href{http://arxiv.org/abs/2003.03921}{ arXiv:2003.03921 [hep-th]}.
	
	\bibitem{Amelino-Camelia:2019cjb}
	G.~Amelino-Camelia, M.~Palmisano, M.~Ronco, and G.~D'Amico, ``{Mixing
		coproducts for theories with particle-dependent relativistic properties},''
	{\em Int. J. Mod. Phys. D} {\bfseries 29} no.~02, (2020) 2050017,
	\href{http://arxiv.org/abs/1910.05997}{ arXiv:1910.05997 [gr-qc]}.
	
	\bibitem{Ballesteros:1999ew}
	A.~Ballesteros, E.~Celeghini, and F.~J. Herranz, ``{Quantum (1+1) extended
		Galilei algebras: from Lie bialgebras to quantum R-matrices and integrable
		systems},'' {\em J. Phys. A} {\bfseries 33} (2000) 3431--3444,
	\href{http://arxiv.org/abs/math/9906094}{ arXiv:math/9906094}.
	
	\bibitem{Amelino-Camelia:2011gae}
	G.~Amelino-Camelia, ``{On the fate of Lorentz symmetry in relative-locality
		momentum spaces},'' {\em Phys. Rev. D} {\bfseries 85} (2012) 084034,
	\href{http://arxiv.org/abs/1110.5081}{ arXiv:1110.5081 [hep-th]}.
	
	\bibitem{Amelino-Camelia:2013sba}
	G.~Amelino-Camelia, G.~Gubitosi, and G.~Palmisano, ``{Pathways to relativistic
		curved momentum spaces: de Sitter case study},'' {\em Int. J. Mod. Phys. D}
	{\bfseries 25} no.~02, (2016) 1650027, \href{http://arxiv.org/abs/1307.7988}{
		arXiv:1307.7988 [gr-qc]}.
	
	\bibitem{Will:2014kxa}
	C.~M. Will, ``{The Confrontation between General Relativity and Experiment},''
	{\em Living Rev. Rel.} {\bfseries 17} (2014) 4,
	\href{http://arxiv.org/abs/1403.7377}{ arXiv:1403.7377 [gr-qc]}.
	
	\bibitem{MICROSCOPE:2022doy}
	{MICROSCOPE} Collaboration, P.~Touboul {\em et~al.}, ``{MICROSCOPE Mission:
		Final Results of the Test of the Equivalence Principle},'' {\em Phys. Rev.
		Lett.} {\bfseries 129} no.~12, (2022) 121102,
	\href{http://arxiv.org/abs/2209.15487}{ arXiv:2209.15487 [gr-qc]}.
	
	\bibitem{Muller:2007zzb}
	J.~Muller and L.~Biskupek, ``{Variations of the gravitational constant from
		lunar laser ranging data},'' {\em Class. Quant. Grav.} {\bfseries 24} (2007)
	4533--4538.
	
	\bibitem{Das:2024xgf}
	S.~Das and S.~Sur, ``{Varying Newton{\textquoteright}s constant: a cure for
		gravitational maladies?},'' {\em Eur. Phys. J. Plus} {\bfseries 139} no.~12,
	(2024) 1114, \href{http://arxiv.org/abs/2411.06489}{ arXiv:2411.06489
		[gr-qc]}.
	
	\bibitem{Singh:2022wyp}
	V.~V. Singh, {\em et~al.}, ``{Equivalence of Active and Passive Gravitational
		Mass Tested with Lunar Laser Ranging},'' {\em Phys. Rev. Lett.} {\bfseries
		131} no.~2, (2023) 021401, \href{http://arxiv.org/abs/2212.09407}{
		arXiv:2212.09407 [gr-qc]}.
	
\end{thebibliography}
\end{document}